# SYMPOSIUM SUMMARY AND OUTLOOK: TWENTY YEARS OF MESON FACTORY PHYSICS

W.C. HAXTON

*Institute for Nuclear Theory, Box 351550, and Department of Physics, Box 351560*
*University of Washington, Seattle, Washington 98195*

The opening of the meson factories twenty years ago provided nuclear physics with new beams, higher momentum transfers, and new opportunities for precision measurements. The resulting changes in nuclear physics were substantial, altering not only the range of physics issues identified with the field but also the manner and size of the collaborations that do nuclear physics. Inspired by the talks of this symposium, I discuss some of the accomplishments as well as some of the goals not yet reached.

## 1 Introduction

It is a great pleasure to take part in this symposium, a celebration of twenty years of new physics from LAMPF and the other meson physics facilities. As in the cases of others who have spoken, LAMPF is a special place for me. My first workshop talk was given here, shortly after I joined T5 as a postdoc early in 1977. It was an occasion I remember fondly, as the few slides I showed stimulated a lively debate between Gerry Brown and Arthur Kerman. The audience seemed to give me some of the credit for helping to provoke their entertaining exchanges. Earlier my thesis, an investigation of inclusive nuclear interactions of neutrinos from stopped pions, had been inspired by Ray Davis's proposal to calibrate his solar neutrino detector at the LAMPF beamstop.

Despite the assigned title for my talk, I hope the audience is not expecting a summary: the excellent talks of the past two days require none. I would instead like to make a few remarks that were inspired by what I have heard at this symposium, but which also connect to my own research. So, in a sense, this is more of a personal tribute to the meson facilities for the intellectual stimulation they have provided and continue to provide.

## 2 Pion-nucleus physics and nuclear structure

The experimental achievements of the meson factory $\pi$-nucleon and $\pi$-nucleus programs were summarized in the excellent talks of Bill Gibbs, Jerry Miller, Bob Redwine, and Anna Hayes. They described a number of substantial achievements: the general accord achieved on low and medium energy $\pi$-p data; credible results on isospin tests; discovery of the relative transparency of the nucleus to pions with E $\sim$ 50 MeV; the successful program to exploit the $\sigma(\pi^+)/\sigma(\pi^-)$ selectivity to probe the isospin character of nuclear transitions; and mapping the isovector monopole resonance.

The Bethe Panel, whose endorsement of the physics potential of meson facilities was crucial to the creation of LAMPF, was quite optimistic about the promise of pions as a nuclear structure probe, especially in their sensitivity to less well understood aspects such as neutron distributions. It is clear that much progress toward this goal was made in the past twenty years: the elementary cross sections were measured quite well; theory produced optical potentials that accurately described elastic scattering; and a substantial body of data was gathered on single and double charge exchange reactions leading to specific nuclear states. Yet, despite this progress, the pion appears not to have furthered our understanding of nuclear structure to the extent envisioned by the Bethe Panel.

While it is easy to attribute this situation to the more complicated reaction mechanism of strongly interacting probes, the parallel nuclear structure studies at electron facilities like Bates also produced puzzling systematics. For example, M1 form factors measured above q $\sim$ 400 MeV [1] tend to remain well above theoretical predictions, even though the electron probe is presumably perturbative and the nuclei in question, such as those in the 1p shell, can be treated with microscopic methods like the shell model.

A common aspect of these Bates measurements and much of the pion physics program at LAMPF is that we are probing the nucleus at momentum transfers large compared to the Fermi momentum. I would like to argue here that some of our difficulties in understanding the nuclear response at high $q^2$ could be attributed to associated shortcuts we make in theory.

While the shell model is often viewed as a phenomenological technique, it is important to remember that it is at least qualitatively motivated by more rigorous many-body theory. Quite generally a problem in a large (or infinite) Hilbert space can be equivalently described in a smaller model space, provided that the interaction used in the model space takes into account the degrees of freedom that were "integrated out [2]." The correct $V_{\text{effective}}$ would reproduce the exact eigenvalues and eigenfunctions (or rather their restrictions to the model space) in the smaller model-space calculation. Now the full calculation of $V_{\text{effective}}$ is a task fully equivalent in complexity to the original A-body problem. The hope in nuclear physics is that an expansion of $V_{\text{effective}}$ in clusters, or equivalently powers of the nuclear density, might rapidly converge [3]. If the convergence is sufficiently rapid, one might then have a tractable strategy for solving the A-body problem.

Unfortunately the shell model has at best a tenuous connection to such rigorous many-body efforts. Most often the shell model potential is simply fitted to spectroscopic data. But lacking a diagrammatic basis for the effective interaction, one has no procedure for making the corresponding corrections that generate the appropriate effective operators from bare operators, a necessary step in evaluating any observable. In effect, the model has lost most of its predictive power. The alternative approach, calculating the effective interaction directly from the NN interaction, is sometimes carried out, but generally only in lowest order, summing the two-nucleon ladder diagram, or G-matrix. Often phenomenological adjustments are made to the results for the purpose of improving agreement with experiment. Matrix element evaluations with consistent treatments of effective interactions and operators are almost never attempted.

Part of the preference for phenomenology may have its roots in the early years of the shell model, when numerical techniques were severely limited by the available computers. For example, convergence problems associated with intruder states [4] were consider a major obstacle; yet modern workstations can easily handle large-basis shell model calculations with many degrees of freedom, such as a full $4\hbar\omega$ calculation of $^{16}$O [5], thereby accounting for all of the principle excitations affecting the low-lying spectrum. Furthermore, powerful new techniques could be applied to the problem of calculating higher order contributions to the effective interaction.

Motivated in part by an INT program organized by Bruce Barrett and James Vary, two of us in Seattle are trying our hand in this old game. The starting point is the Bloch-Horowitz equation [6] for the effective Hamiltonian

$$H_{\text{effective}} = (T+V) + (T+V)\frac{1}{E-Q(T+V)}Q(T+V) \qquad (1)$$

where T is the relative kinetic energy operator, V is the NN potential, and Q projects onto the excluded, high-momentum space. The hope in nuclear physics, which dates back to arguments made by Brueckner [3], is that the second term in this equation might be expandable in a series in the nuclear density, corresponding to the ladder diagrams for two, three, four, ..., nucleons interacting at short range.

The Bloch-Horowitz equation looks very difficult to solve because the energy E appearing in the Green's function is the exact eigenvalue. Thus, for each desired state, one would have to solve this equation selfconsistently, presumably iterating until convergence is achieved. If one had to reevaluate the effective interaction for each trial E, the algorithm would be clearly quite inefficient.

However the Lanczos algorithm, quite commonly used to find the extremum eigenvalues and eigenvectors of very large matrices, has a less well known application to linear operator inversion [7] that circumvents this difficulty. The Lanczos algorithm solves the moments problem for the distribution of $H$ over its eigenvalues, determining the $2n-1$ moments of $H$ after $n$ iterations. These moments can be determined from the tridiagonal Lanczos matrix. From knowledge of these moments one can calculate the effects of operators like the Green's function, which sums over the full spectrum. One finds

$$\frac{1}{E-Q(T+V)}|v_1\rangle \simeq g_1(E)|v_1\rangle + ... + g_n(E)|v_n\rangle \qquad (2)$$

where the $|v_i\rangle$ are the Lanczos vectors and the $g_i$ are continued fractions determined by E and by the entries in the Lanczos matrix. If one chooses the starting Lanczos vector $|v_1\rangle$ to be $Q(T+V)|I\rangle$ where $|I\rangle$ is a state in the model (included) space, this expansion can be used to determine the matrix elements of $H_{\text{effective}}$. Because the energy $E$ is merely a parameter in the continued fractions, once one carries out the Lanczos expansion, the effective interaction is known as a function of E. No additional computer time is required to iterate on E. As the Lanczos expansion itself is quite efficient in determining the Green's function, typically converging in about 50 iterations, this becomes a very powerful method for solving the Bloch-Horowitz equation.

We believe it may be possible to calculate the effective interaction in a cluster expansion through the three- and, possibly, four-nucleon ladder, thereby correcting the usually two-nucleon G-matrix through second order in the nuclear density. It is an open question whether this is sufficient to achieve convergence.

This sketch of how one might attempt to generate $H_{\text{effective}}$ given modern workstations and efficient algorithms, could be repeated for effective operators. This is the reason for including the discussion here. Returning to the example of exclusive nuclear form factors measured at Bates, one can envision operating on a model-space wave function by the magnetic operator $M1(q^2)$ transferring, for example, twice the Fermi momentum. The resulting amplitude has a very small overlap with the model space relative-coordinates wave function describing the final state, as this wave function contains only long wavelength components. The scattered nucleon must interact with one or more neighboring nucleons, sharing the momentum, before the overlap becomes appreciable. That is, it is the effective contributions to the operator that should dominate the response. Therefore, if we fail to correct both our interactions and operators consistently, we will lack the basic ingredients for a realistic calculation of nuclear responses at high $q^2$. And it is clear, as $q^2$ increases, we will need to include increasingly complicated correlations in our effective operator to achieve a realistic result. The arguments here are very much like those that lead to quark counting rules for form factors at asymptotically large $q^2$.

I believe our lack of effort on this problem - calculating consistent effective operators and interactions from realistic NN interactions - may have limited the meson factory (and other medium energy facility) impact on our understanding of nuclear structure.

## 3 Neutron Physics

Stuart Freedman, David Bowman, John Browne, and Steve Vigdor described many of the elegant symmetry tests that have been done with low and medium energy neutrons, as well as the prospects for future experiments at LANSCE and LISS. The results include remarkably precise limits on CP- and T-violation from neutron electric dipole moment and D-coefficient $\beta$ decay measurements, as well as the demonstration that parity-nonconserving resonance mixing can probe chaotic properties of the nucleus.

One interesting issue not discussed is the relation of neutron and atomic electric dipole moments measurements. The neutron edm limit from Genoble [8] and Gatchina [9]

$$|d_n| \lesssim 8 \cdot 10^{-26} \text{e cm} \qquad (3)$$

can be compared to the Seattle group's [10] limit on the mercury edm

$$|d(^{199}\text{Hg})| \lesssim 1.3 \cdot 10^{-27} \text{e cm} \qquad (4)$$

As the $^{199}$Hg atom is probed with electric fields of about $10^4$ volts/cm, this corresponds to a change in the ground state energy of the atom of about $10^{-23}$eV, which is approximately the interaction energy of two electrons separated by one light-month! This certainly illustrates the extraordinary precision of the limits above.

Although the $^{199}$Hg edm limit is about sixty times smaller than that of the neutron, deciding which is the more stringent test of underlying sources of CP violation is quite complicated. The nuclear edm of an odd-A nucleus is generated both by the edm of the unpaired valence nucleon and by a nuclear polarizability: a CP-odd NN interaction produces a parity admixture in the ground state that allows absorption of a C1 (J=1 multipole of the Coulomb interaction) photon. The latter is generally the dominant contribution to the nuclear edm [11]. Depending on the nature of the underlying CP-violating Lagrangian, the entire nucleus may contribute to the polarizing nuclear mean field, or the effects may only arise from the unpaired neutrons. This means that the question of relative sensitivity of neutron vs. atomic limits is quite dependent on the CP violation model.

Further complicating the comparison is the Schiff shielding [12] of the atom: for a point nucleus in a neutral atom, the atomic interaction energy linear in the edm vanishes. The atomic cloud sags in the applied field to cancel the net field at the nucleus. (If this did not happen, the nucleus of a neutral atom would be accelerated by the external field.) In a real atom a nonzero interaction is found because of the nuclear finite size, hyperfine effects, and relativistic corrections. The residual effects grow rapidly with Z, making heavy atoms the favorites of experimentalists.

The net result - nuclear coherence in the polarizability, the loss of sensitivity due to electronic shielding, and the factor of 60 smaller $^{199}$Hg edm limit - is comparable sensitivity to CP-violation sources that are coherent over the entire nucleus (such as the KM phase), and a modest advantage in favor of the neutron edm when the coherence is only over valence neutrons (such as in the case of the QCD $\theta$ parameter) [13].

This situation may soon change. Recent improvements in the nuclear measurements have been very rapid, outpacing the corresponding improvements in the neutron edm limit. Furthermore, there are several interesting nuclei that should have greatly enhanced C1 polarizabilities, in one case by an estimated factor of $10^4$ [11]. While the best cases are not attractive experimental candidates, it is quite possible that this could change as new techniques (e.g., trapping experiments) [14] are developed. It is exciting that such advances might allow experimentalists to probe edms near the $10^{-31}$ e cm range predicted to arise from the standard model KM phase.

## 4 Neutrinos

Peter Nemethy, Peter Rosen, and Michael Shaevitz described the marvelous LAMPF contributions to neutrino physics (E31, E225, E645, and LSND). This continues an even longer tradition at Los Alamos that includes the discovery of the neutrino by Cowan and Reines, for which Reines won the Nobel Prize this year. The LAMPF experimental program provided crucial tests of additive vs. multiplicative conservation laws for lepton number and constrained the sign of the charge current - neutral current interference in $(\nu, e)$ scattering. Most exciting is the current experiment, LSND, with 52 events attributed to neutrino oscillations [15].

One reason for the intense interest in neutrino physics has to do with the puzzling pattern of masses in the standard model. If this model is part of some grander, more unified scheme, one would anticipate finding larger multiplets in that extended scheme, perhaps uniting particles by generation, e.g., $(e, \nu_e, u, d)$. But as these particles presumably would have similar couplings to the fields generating masses, the much lighter mass of the $\nu_e$ then seems a puzzle. However $\nu$s are distinguished from the other particles in the multiplet because they have no additive quantum numbers that must change sign under particle-antiparticle conjugation. Thus in addition to the Dirac masses of the other fermions, neutrinos can have Majorana masses. In the popular seesaw mechanism, the introduction of a large right-handed

Majorana mass $M_R$ leads to light and heavy neutrinos

$$m_L \sim m_D \left[\frac{m_D}{M_R}\right] \tag{5}$$

$$m_H \sim M_R \tag{6}$$

with the familiar $\nu_e$ primarily identified with the light eigenstate $m_L$. Thus one sees that small neutrino masses are explained by a parameter, $m_D/M_R$, reflecting familiar and new mass scales. It follows that the ability to probe small neutrino masses becomes a search for very high scales $M_R$ characterizing new physics. If we take $M_R \sim 10^{16}$ GeV, a typical grand unification mass, and $m_D \sim m_{top\ quark}$, one finds a third generation neutrino mass of $m_{\nu_\tau} \sim 10^{-3}$ eV, a number accessible to neutrino oscillation experiments.

I think it would be a great disappointment if the LAMPF neutrino program were to come to an premature end: the beamstop remains our most intense terrestrial source of low-energy $\nu_e$s, $\nu_\mu$s, and $\bar{\nu}_\mu$s. In the near term, the moving of LSND ought to be given very high priority: the associated cost of $5M is quite modest on the scale of the average neutrino experiment. The current results of that experiment make it imperative that every step be taken to test the results. On the longer term, I feel it is very important to the nation's nuclear physics program to keep a neutrino presence on the mesa. The neutron physics material science program at LANSCE calls for several possible upgrades. There is no reason that neutron and neutrino programs should not coexist: both benefit from intensity upgrades. I find it surprising, in view of the marvelous neutrino physics legacy of Los Alamos, that the community has not fought harder to keep the LAMPF/LANSCE neutrino program alive.

Before closing, I would like to mention some other recent work relevant to LSND where neutron and neutrino physics are intertwined. In nuclear astrophysics it is known that approximately half of the elements heavier than Fe must be synthesized in the rapid-neutron-capture process, or $r$-process. The conditions for this process are very exotic, requiring explosive neutron fluxes on the order of $10^{28}/\text{cm}^2$s lasting for $\sim 1$ s [10]. Under these conditions neutron capture becomes faster than $\beta$ decay, so that the synthesis follows a path through very neutron rich nuclei defined by $(n,\gamma) \leftrightarrow (\gamma,n)$ equilibrium. When the neutron flux diminishes, these isotopes then decay back to the valley of stability, forming the $r$-process nuclei we identify in nature.

The question of the site of the $r$-process has been debated for almost 40 years. One of the most popular suggestions has been in the neutronized matter near the mass cut that is blown off the protoneutron star during a core-collapse supernova. This is the site relevant to LSND.

The protoneutron star cools by emitting neutrinos of all flavors. While the neutrinos remain in flavor equilibrium (due to neutral current interactions) throughtout most of their random walk out of the dense neutron star core, as they reach the neutrino sphere (the radius of last scattering) at $\sim 10^{12}$ g/cm$^3$, some differences arise. The electron-flavor neutrinos are more strongly coupled to the matter because of the charged current interactions

$$\nu_e + n \to p + e^- \tag{7}$$

$$\bar{\nu}_e + p \to n + e^+ \tag{8}$$

As a result, the distributions of heavy-flavor neutrinos are characterized by temperatures $T_{\nu_\mu} \sim T_{\nu_\tau} \sim 8$ MeV, while $T_{\bar{\nu}_e} \sim 5$ MeV and $T_{\nu_e} \sim 4$ MeV. (The latter difference arises because the matter is neutron rich.)

Now if neutrinos have a cosmologically interesting mass ($m_{\nu_\mu}$ or $m_{\nu_\tau} \sim 2-50$ eV) and even very modest mixing angles ($\sin^2 2\theta \gtrsim 10^{-5}$), neutrinos can destroy the supernova $r$-process. Matter-enhanced oscillations between, for example, $\nu_e$s and $\nu_\tau$s cause a temperature reversal between these species. As Qian, Fuller, and collaborators have discussed [17], the hot $\nu_e$s then have substantially increased charge current interactions, driving the matter above the neutron star proton rich. The $r$-process could no longer take place.

Much of the oscillation parameter space suggested by LSND would then be eliminated by the requirement of a successful $r$-process. Of course, there appears to be an escape clause: perhaps the hot bubble above the protoneutron star is not the site of the $r$-process. As many other sites have been suggested, this is plausible and would invalidate any constraints on LSND.

Recently Langanke, Qian, Vogel and I [18] explored another aspect of this $r$-process site, the fact that the neutrino flux would continue for several seconds after the $r$-process has been completed (which occurs when the matter temperature drops to $\sim 10^9$K). These late neutrinos would occasionally interact, by charged or neutral currents, with the $r$-process products. Typical reactions produce highly excited nuclei that decay by emitting several neutrons. Thus this neutrino postprocessing could alter the $r$-process abundance distribution.

When we studied the postprocessing in detail, we discovered that there exist 8 rare nuclei (Te, W, and Re isotopes) that are extremely sensitivity to the neutrino postprocessing. These isotopes live in the abundance valleys immediately below the $r$-process abundance peaks associated with the closed neutron shells at N = 50 and 82. Thus a few neutrino interactions with the isotopes in the abundance peaks can produce appreciable quantities of Te, W, and Re by spallation. When this possibility was explored in detailed, it turned out that the abundances of all eight isotopes could be entirely accounted for by neutrino spallation: the agreement with the known abundances is excellent ($\sim 1\sigma$). Furthermore the neutrino fluence required to produce these nuclei by spallation is in good accord with supernova models of the $r$-process. Some results are illustrated in Fig. 1.

This result strongly argues that these isotopes are neutrino postprocessing products, and therefore that the $r$-process does take place in the intense neutrino flux in the atmosphere above a protoneutron star. It follows that the $r$-process limit on neutrino oscillations [17] then must be taken as a very serious additional constraint on the same parameter space probed by LSND.

## 5 Closing

Let me close this session by saying "Thanks!" to
* LAMPF, for 20 great years
* Louis Rosen, for making it happen
* our workshop hosts, Peter Barnes and Vernon Hughes, for allowing us to come together one last time
* the organizing committee Cy Hoffman (chair), Ben Gibson, Bill Louis, Susan Ramsey,

Susan Seestrom, and Dave Vieira, for putting together a wonderful two days.

This work was supported in part by the U.S. Department of Energy.

**Figure Captions**

Figure 1: The solid line represents the $r$-process distribution that would be produced if, after freezeout, there is no further processing by neutrinos. The dotted line represents the changes due to the postprocessing by neutrinos (corresponding to a total neutrino fluence of $0.015 \cdot 10^{47}$ erg/km$^2$). The fluence was chosen so that isotopes in the window A = 183-187 could be produced entirely by neutrinos. The quality of the resulting fit to abundances in this window strongly suggests that the $r$-process occurs in an intense neutrino flux. See Ref. [18] for more details.

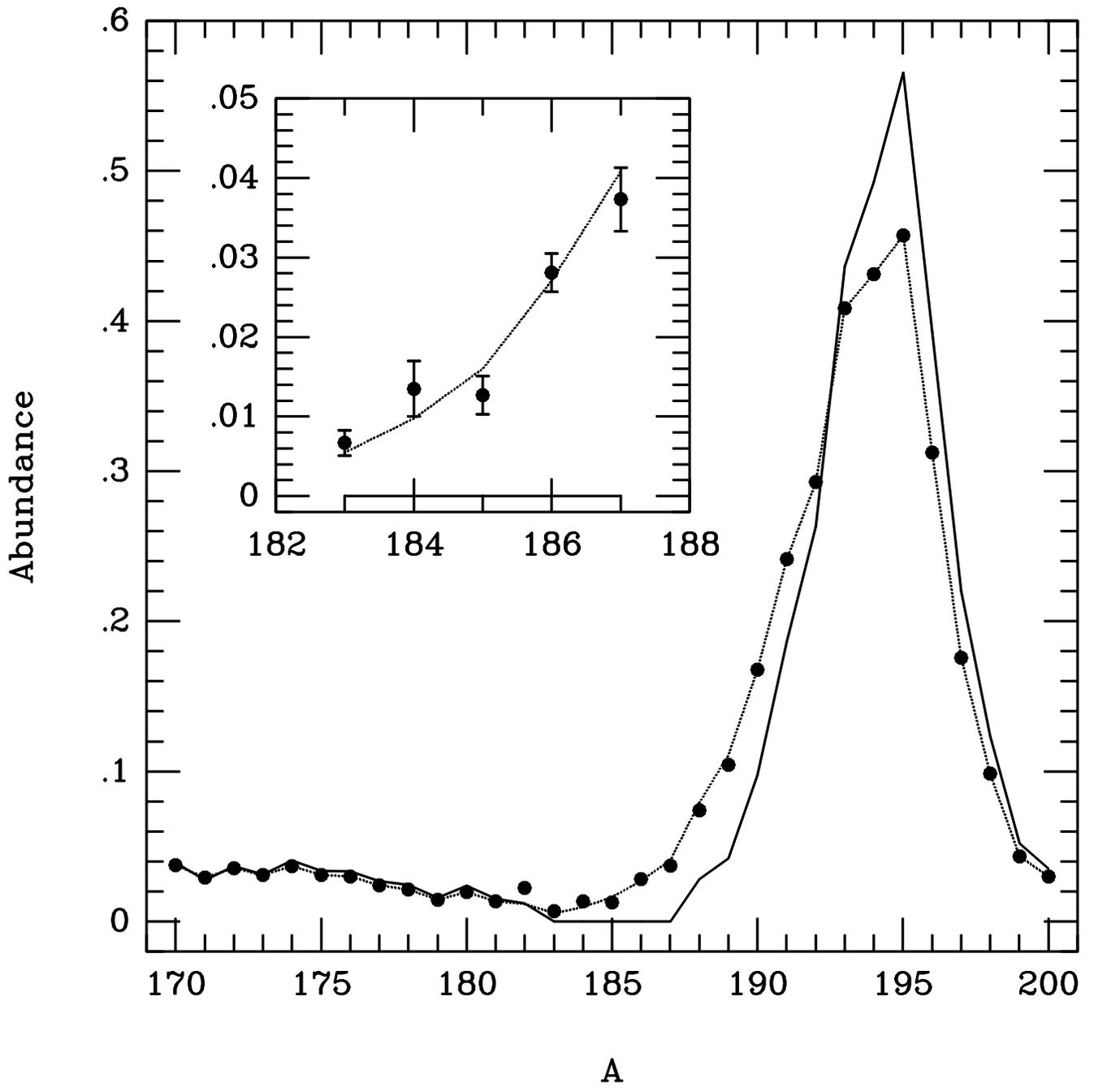